\documentclass[hidelinks, reprint, nofootinbib, amsmath, amssymb, aps, longbibliography]{revtex4-2}

\usepackage{graphicx}
\usepackage{dcolumn}
\usepackage{bm}
\usepackage{hyperref}

\usepackage{natbib}

\begin{document}

\title{Possibility of Small Electron States\\\vspace*{6 pt}\normalsize{arXiv v.2}}
\author{\vspace*{-6 pt}[forthcoming in \textit{Physical Review A}]\\\vspace*{6 pt}Charles T. Sebens}
\email{csebens@caltech.edu}
\affiliation{Division of the Humanities and Social Sciences, California Institute of Technology, Pasadena, CA, USA}
\date{November 25, 2020}

\begin{abstract}
Some authors have claimed that there exists a minimum size (on the order of the Compton radius) for electron states composed entirely of positive-frequency solutions to the free Dirac equation.  Other authors have put forward counterexamples to such claims.  This article asks how the counterexamples of A. J. Bracken and G. F. Melloy [J. Phys. A. \textbf{32}, 6127 (1999)] bypass two arguments against their possibility.  The first is an old argument that, because of the prohibition on faster-than-light motion, the electron must be larger than a certain minimum size if it is to have the correct angular momentum and magnetic moment.  This challenge can be addressed by analyzing the flow of energy and charge for the counterexample states.  The second argument is an explicit proof (presented in C.-P. Chuu \emph{et al.}, [Solid State Commun. \textbf{150}, 533 (2010)]) that there is a minimum size for purely positive-frequency electron states.  This proof hinges on the assumption of a small spread in momentum space, which is violated by the counterexamples that have been put forward.
\end{abstract}

\maketitle

\section{Introduction}\label{intro}

The free Dirac equation has both positive-frequency solutions (associated with electrons) and negative-frequency solutions\footnote{The negative-frequency solutions are often called negative-energy solutions, but I avoid that terminology because negative-frequency solutions do not have to be attributed negative energy \cite{positrons}.} (associated with positrons).  Using only the positive-frequency modes, many physicists have claimed there is a minimum size for the electron states that can be constructed, on the order of the Compton radius (or Compton wavelength) $\frac{\hbar}{mc}$.  However, Bracken, Flohr, and Melloy \cite{bracken1999, melloy2002, bracken2005} have put forward counterexamples to such claims, showing that there exist arbitrarily small electron states composed entirely of positive-frequency modes.

In this article, I analyze these counterexamples and explain how they avoid two arguments against the possibility of small electron states that appear in Refs.\ \cite{chuu2007, chuu2010, chuu2010thesis}.  The first argument dates back to Kronig and Lorentz around 1925:\footnote{See Refs.\ [\onlinecite{kronig1926}, \onlinecite[pg.\ 47]{uhlenbeck}, \onlinecite[pg.\ 35]{tomonaga1997}].} given that no parts of the electron move faster than the speed of light, the electron must be larger than a certain minimum size for it to be possible that the electron's rotation generates the correct magnetic moment and angular momentum.  The second argument is a proof that there is a minimum size for electron states composed of positive-frequency solutions to the Dirac equation.  The first argument can be addressed by studying the densities of charge, energy, current, and momentum for small electron states.  The magnetic moment of a small electron state can be explained without faster-than-light rotation of charge because small electron states have weak magnetic moments.  The angular momentum of a small electron state can be explained without faster-than-light rotation of energy because small electron states have large energies.  The second argument does not manage to rule out small electron states because the proof relies on an assumption that the momentum space representation of the electron's state is tightly peaked, which is not true for the counterexample states in \cite{bracken1999, melloy2002, bracken2005}.

The free Dirac equation,\footnote{This way of writing the Dirac equation employs the Dirac gamma matrices \cite[ch.\ 2]{bjorkendrell}.}
\begin{equation}
i \hbar \frac{\partial \psi}{\partial t}=\big(-i \hbar c \: \gamma^0 \vec{\gamma}\cdot\vec{\nabla} + \gamma^0 m c^2 \big)\psi
\ ,
\label{dirac}
\end{equation}
is commonly understood as part of a relativistic quantum mechanics for the electron, describing the dynamics of a four-component complex-valued single-particle wave function $\psi$.  The probability density for position measurements of the electron is $\psi^{\dagger} \psi$ and the probability current is $c\psi^{\dagger} \gamma^0 \vec{\gamma} \psi$ (or in different notation, $c\psi^{\dagger} \vec{\alpha} \psi$).  Often, the electron wave function is described as possessing densities of charge and charge current proportional to the densities of probability and probability current,
\begin{align}
\rho^q&=-e \psi^\dagger \psi
\label{diracchargedensity}
\\
\vec{J}&=-e c \psi^\dagger \gamma^0 \vec{\gamma} \psi
\ ,
\label{diraccurrentdensity}
\end{align}
where $-e$ is the charge of the electron and the $q$ superscript indicates that $\rho^q$ is a density of charge.  This is the way that $\psi$ is understood in Refs.\ \cite{bracken1999, chuu2007, chuu2010, chuu2010thesis}.

Alternatively, one might interpret the Dirac equation as part of a classical theory of the Dirac field (which is of interest because it yields the correct quantum field theory of electrons and positrons after field quantization).  On this interpretation, $\psi$ is a four-component complex-valued classical field.  As above, we can take the charge density and current density to be given by \eqref{diracchargedensity} and \eqref{diraccurrentdensity} (though $\psi^{\dagger} \psi$ would not be a probability density in this classical theory).\footnote{In Ref.\ \cite{positrons} I argued that we should actually flip the sign of the charge density and charge current density for negative-frequency modes.  As we are not analyzing negative-frequency modes here, we can put this issue aside.}  I have defended the fruitfulness of this classical field interpretation of $\psi$ in Refs.\ \cite{howelectronsspin, positrons}, where I made use of what \textcite{bracken2005} describe as a false but widely believed ``folk theorem'': that there is a minimum size (on the order of the Compton radius) for electron states composed of positive-frequency modes.  In this article, I recant my previous claims of a minimum size for the distribution of charge in states of the Dirac field representing a single electron.  In unusual circumstances, smaller electron states are possible.  With this correction made, the picture of electron spin in Refs.\ \cite{howelectronsspin, positrons, electronsspinmeasurement} remains viable.

For our purposes here, we need not choose between the two interpretations of $\psi$ presented above.  The analysis applies to both.  One can read the body of this paper as a discussion of either the possibility of compact electron wave functions in single-particle relativistic quantum mechanics or the possibility of compact Dirac field configurations in classical Dirac field theory.  In the conclusion, these two options are separated to briefly discuss the connection to quantum field theory.

Let us focus on states where the electron's charge is centered at the origin,
\begin{equation}
\langle \vec{x}\, \rangle =\frac{1}{-e} \int \! d^3 x\; \rho^q \vec{x} = 0
\ .
\end{equation}
When evaluating whether there is a minimum size for electron states, we can use the mean square charge radius as the relevant measure of size,
\begin{equation}
\langle |\vec{x}|^2 \rangle =\frac{1}{-e} \int \! d^3 x\; \rho^q |\vec{x}|^2
\ .
\label{chargeradius}
\end{equation}
This is a charge-weighted average of the squared distance from the center of charge.  The mean square charge radius can of course be small even if the electron state has tails of decreasing charge density stretching off to infinity.\footnote{For discussion of tails in electron wave functions, see Refs.\ \cite{bakke1982, thaller1984, hegerfeldt1989, hegerfeldt1998, bracken1999, bracken2005}.}

\section{Claims of a Minimum Size}\label{CMSsection}

A number of prominent textbooks from the past century include claims of a minimum size for electron states built from positive-frequency modes.  After using the Dirac sea to motivate a focus on positive-frequency modes, \textcite{heitler} writes:
\begin{quote}
``Consequently, the negative energy [negative-frequency] contributions to the wave function must be eliminated and the electron cannot be a wave packet of infinitely small size but must have a finite extension (of the order $\frac{\hbar}{mc}$, as one easily finds).'' \cite[pg.\ 299]{heitler}
\end{quote}
But no proof is given.  In the previous edition of this text, \textcite[pg.\ 193]{heitler2} writes that ``it has only a limited meaning to speak of'' electron wave packets smaller than the Compton radius because they include modes with corresponding energies that are large enough to create electron-positron pairs.  This observation about the presence of high energy modes is relevant to determining how well the behavior of electron states in a particular physical theory (without pair creation) approximates the behavior of electron states in a deeper theory (with pair creation).  However, this observation does not tell us what states are possible within a given physical theory.  For the context under consideration here, note that pair creation will not occur for purely positive-frequency states evolving under the free Dirac equation.  For theories that include pair creation, note that claims about small electron states generating electron-positron pairs are claims about the dynamics (which presuppose that such states are possible at a moment).

\textcite[ch.\ 3]{bjorkendrell} show that for a certain kind of Gaussian wave packet with a variable width (composed of both positive- and negative-frequency modes), negative-frequency modes only become significant if the width is near or below the Compton radius.  They then write:
\begin{quote}
``This result can be equally well inferred on dimensional groups [\emph{sic}] using $\Delta p \Delta x \sim \hbar$ without reference to the particular gaussian shape.  In discussing problems and interactions where the electron is `spread out' over distances large compared with its Compton wavelength, we may simply ignore the existence of the uninterpreted negative-energy solutions and hope to obtain physically sensible and accurate results.  This will not work, however, in situations which find electrons localized to distances comparable with $\frac{\hbar}{mc}$.  The negative-frequency amplitudes will then be appreciable...'' \cite[pg.\ 40]{bjorkendrell}
\end{quote}
But the dimensional argument that they sketch would not apply to states that are formed entirely of positive-frequency modes (where there would be no possibility for negative-frequency modes to grow more significant as we consider smaller sizes).\footnote{Citing \textcite{bjorkendrell}, \textcite[pg.\ 635]{barut1968} write that a restriction to positive-frequency modes ``cuts the Hilbert space of states in half and brings about an important difference between position properties of Schr\"{o}dinger and Dirac particles: A Schr\"{o}dinger particle can be localized within as small a volume as we please, while a Dirac particle cannot be localized to within less than, roughly speaking, its Compton wavelength.''}

In a passage quoted in Refs.\ \cite{bakke1973, almeida1984}, \textcite{sakuraiAQM} writes:
\begin{quote}
``We have seen that a well-localized state contains, in general, plane-wave components of negative energies.  Conversely, we may ask how well localized a state will be which we can form using only plane-wave components of positive energies.  A rather careful analysis by T. D. Newton and E. P. Wigner indicates that the best localized state we can construct in this way is one in which the characteristic linear dimension of the wave packet is $\sim\!\frac{\hbar}{mc}$ but no smaller.'' \cite[pg.\ 119]{sakuraiAQM}
\end{quote}
But \textcite{newton1949} do not directly address this question and their results do not entail a minimum size for such states \cite{bracken1999, bracken2005}.  In discussing the relativistic quantum mechanics of spin-$0$ particles, \textcite[sec.\ 2.G]{feshbach1958} also cite \textcite{newton1949} and claim that there is a minimum size for wave packets built of positive-frequency modes.  However, I see no explicit argument for this claim.  Feshbach and Villars note that the eigenstates of a specified ``mean position operator''\footnote{When the position operator acts on a purely positive-frequency state, it will yield a superposition of positive- and negative-frequency modes.  They call the operator that yields only the positive-frequency modes the ``even part'' of the position operator or the ``mean position operator'' (following Ref.\ \cite[pg.\ 32]{foldy1950}).} contain only positive-frequency modes and are each spread out over the Compton radius, but do not show that all states containing only positive-frequency modes are spread out at least this widely.  Later in the paper, \textcite[sec.\ 3.E]{feshbach1958} move to spin-$1/2$ particles and write, ``As in the spin zero case, here again it is impossible to construct purely positive packets that are localized any better than $\frac{\hbar}{mc}$.''

\section{Small Electron States}\label{SESsection}

There is an old argument against the possibility of small electron states that appeals to the general prohibition on faster-than-light motion in relativistic theories.  One version of this argument begins with the standard\footnote{In the context of the free Dirac equation, \eqref{dirac}, the magnetic moment of the electron is given by the Bohr magneton, \eqref{bohrmagneton}.  Quantum electrodynamics accurately predicts the small difference between the actual magnetic moment of the electron and the Bohr magneton.} magnetic moment for the electron (the Bohr magneton),\footnote{Here and throughout I adopt Gaussian cgs units.}
\begin{equation}
|\vec{m}|=\frac{e\hbar}{2 m c}
\ .
\label{bohrmagneton}
\end{equation}
From the relation between current density and magnetic moment,
\begin{equation}
\vec{m}=\frac{1}{2c}\int \! d^3 x \left(\vec{x} \times \vec{J}\,\right)
\ ,
\label{totalmagneticmoment}
\end{equation}
and the assumption that this current density is the charge density $\rho^q$ times some velocity field $\vec{v}^{\,q}$ describing how each bit of electron charge is moving,
\begin{equation}
\vec{v}^{\,q}=\frac{\vec{J}}{\rho^q}
\ ,
\label{chargevelocity}
\end{equation}
we get the consequence that the electron's charge cannot be distributed too compactly.  It must be spread out over (approximately) the Compton radius if we are to avoid faster-than-light velocities.\footnote{See Ref.\ \cite[sec.\ 2]{howelectronsspin}.}  Each bit of charge makes a contribution to the magnetic moment that is larger the faster it is moving and the farther out it is from the axis of rotation.  The smaller the electron is, the faster it must rotate to maintain the magnetic moment in \eqref{bohrmagneton}.  With a limit on the maximum speed comes a limit on the minimum size.  A second version of this argument reaches the same conclusion by appealing to the known angular momentum associated with electron spin, $\frac{\hbar}{2}$, and a prohibition on the electron's energy (or relativistic mass) moving faster than the speed of light.  Let us start by analyzing the flow of charge and the first version of the argument, then move on to analyze the flow of energy and the second version of the argument.

From the expressions for charge density and current density in \eqref{diracchargedensity} and \eqref{diraccurrentdensity} one can prove that: for any state $\psi$, the velocity of charge flow in \eqref{chargevelocity} cannot exceed the speed of light [\onlinecite[sec.\ 2b]{takabayasi1957}; \onlinecite[sec.\ 10.4]{bohmhiley}; \onlinecite[sec.\ 12.2]{holland}].  Despite this speed limit, the small electron states in Refs.\ \cite{bracken1999, melloy2002, bracken2005} are in fact possible because the magnetic moment calculated from \eqref{diraccurrentdensity} and \eqref{totalmagneticmoment} can be weaker than \eqref{bohrmagneton}.  When we consider progressively smaller electron states below, we see that the magnetic moment generated by the flow of the electron's charge approaches $0$.  Thus, the above argument fails to show that there exists a minimum size for electron states formed by superposing positive-frequency solutions to the Dirac equation.  However, it does show that there is a minimum size for states that actually have the magnetic moment in \eqref{bohrmagneton}.

\begin{figure*}[tb]
\includegraphics[width=15 cm]{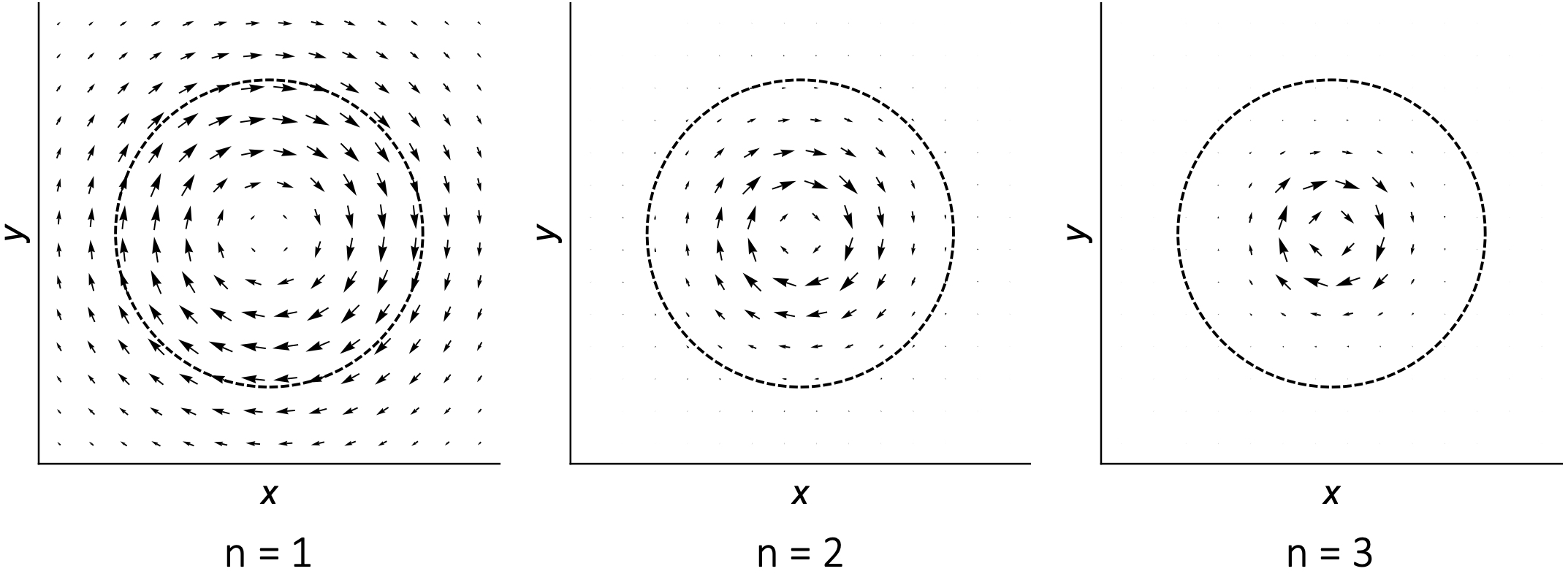}
\caption{These three plots show the current density at $z$=0 of \eqref{generalstate} for three values of $n$.  The plots include a dashed circle with a radius of $\frac{\hbar}{mc}$ (the Compton radius).  For each electron state, the current circles the $z$ axis and generates a magnetic moment pointing in the minus $z$ direction.  As $n$ increases, the charge distribution shrinks and as it does the region where the current is significant gets smaller (decreasing well below the Compton radius).  If the same scale were used for the arrows in all three plots, the arrows in the second plot would be much larger than those in the first and the arrows in the third plot would be larger still (because the peak current density is higher for larger $n$).  When you take into account both the decrease in electron size and the increase in peak current density, the total magnetic moment, \eqref{totalmagneticmoment}, decreases as $n$ increases.}
\label{currents}
\end{figure*}

Bracken \emph{et al.}\ \cite{bracken1999, melloy2002, bracken2005} have proposed counterexamples to the ``folk theorem'' that there is a minimum size for electron states composed entirely of positive-frequency modes.  Earlier, \textcite{bakke1973} proposed counterexamples in the context of the one-dimensional Dirac equation.  Here we analyze flow of charge for a set of states discussed in Refs.\ \cite{bracken1999, melloy2002, bracken2005}:
\begin{equation}
\psi_n(\vec{x}\,)=\int \! \frac{d^3 p}{(2\pi\hbar)^{3/2}} \,  f_n(\vec{p}\,) \, u_1(\vec{p}\,) \,e^{\frac{i}{\hbar}\vec{p}\cdot\vec{x}}
\ ,
\label{generalstate}
\end{equation}
where $n$ can be varied.  These states are written as a superposition of positive-frequency solutions to the Dirac equation (at a moment)\footnote{As our focus is on what is possible at a moment, the time evolution of these states is not included in \eqref{generalstate} (see Ref.\ \cite{bracken2005}).} for each momentum $\vec{p}$ using the standard spinors,\footnote{See Refs.\ [\onlinecite[pg.\ 510]{melloy2002}; \onlinecite[eq.\ 3]{chuu2010}; \onlinecite[eq.\ 2.13]{chuu2010thesis}; \onlinecite[ch.\ 3]{bjorkendrell}; \onlinecite[pg.\ 84]{schweberQFT}].  One could also include the spinors $u_2(\vec{p}\,)$ in forming purely positive-frequency electron states.  However, they are not needed to show that arbitrarily small electron states are possible.}
\begin{equation}
u_1(\vec{p}\,) = \frac{1}{\sqrt{2 E(\vec{p}\,)(E(\vec{p}\,)+m c^2)}}\left(
\begin{matrix}
E(\vec{p}\,)+m c^2 \\
0 \\
p_z c \\
p_x c + i p_y c
\end{matrix}\right)
\ ,
\end{equation}
where $E(\vec{p}\,)=\sqrt{|\vec{p}\,|^2 c^2 + m^2 c^4}$ (the relativistic energy of a point particle with rest mass $m$ and momentum $\vec{p}\,$).  For the function $f_n(\vec{p}\,)$ assigning amplitudes to each of these spinors, let us adopt the simple choice of an appropriately normalized Gaussian distribution in momentum space centered around the origin,
\begin{equation}
f_n(\vec{p}\,)=\frac{1}{(n m c)^{3/2}\pi^{3/4}}e^{\frac{-|\vec{p}\,|^2}{2 n^2 m^2 c^2}}
\ .
\label{momentumspread}
\end{equation}
A small value for $n$ yields a state \eqref{generalstate} with a narrow spread in momentum space and a wide spread in position space.  Increasing the value of $n$, you can generate a state with an arbitrarily wide spread in momentum space and an arbitrarily narrow spread in position space.  Thus, the mean square charge radius, \eqref{chargeradius}, can be made arbitrarily small.  To see this shrinking, numerical integration can be used to perform the Fourier transformation in \eqref{generalstate} and calculate the charge density, \eqref{diracchargedensity} \cite[fig.\ 1]{bracken1999}.

I focus here on understanding the above electron states at a moment and do not examine their time evolution.  \textcite{bracken2005} have shown that for large $n$ the states rapidly expand.  Electron states do not stay smaller than the Compton radius for long.  For further discussion of the time evolution of such states, see Refs.\ \cite{bakke1973, almeida1984, decastro1991}.

\begin{figure}[t]
\includegraphics[width=8.6 cm]{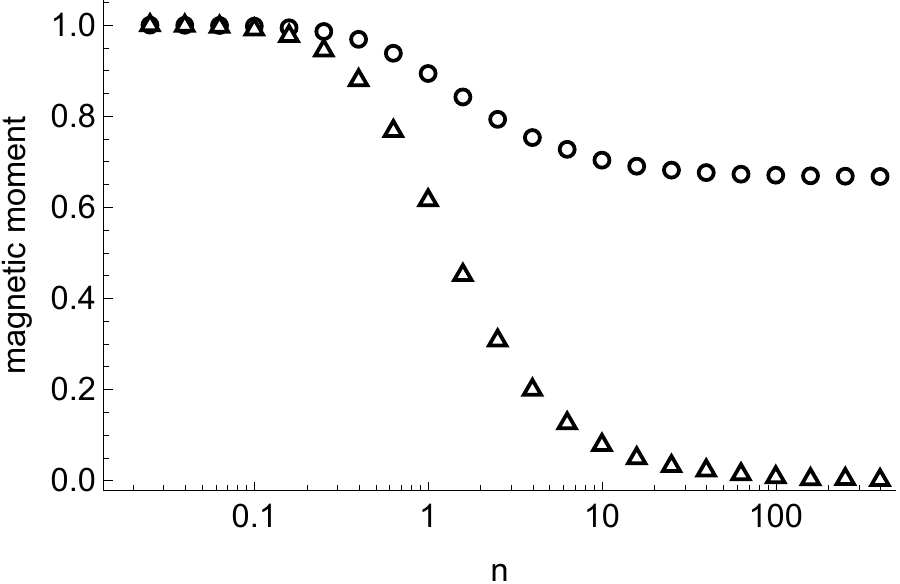}
\caption{This plot shows the total magnetic moment (triangles) and the spin contribution to the magnetic moment (circles) for different values of $n$, using a log scale for $n$.  The magnetic moment is given as a fraction of the Bohr magneton, \eqref{bohrmagneton}.  For small values of $n$, both the total magnetic moment and the spin contribution approach the Bohr magneton.  For large values of $n$, the total magnetic moment approaches $0$ and the spin contribution approaches two-thirds of its original value.}
\label{moments}
\end{figure}

Even though the charge distributions for the electron states in \eqref{generalstate} shrink as $n$ increases, there is no need for faster-than-light rotation.  From \eqref{diraccurrentdensity} and \eqref{generalstate} we can (via numerical integration) calculate the current density describing the flow of charge at any point in space.  These current densities are plotted for three increasing values of $n$ in figure \ref{currents}.  From \eqref{diraccurrentdensity}, \eqref{totalmagneticmoment}, and \eqref{generalstate} we can calculate (again using numerical integration\footnote{To calculate the $z$ component of the total magnetic moment \eqref{totalmagneticmoment}, one needs to integrate $x J_y - y J_x$ over all of space.  For the $x J_y$ term, there are nine integrals to be performed (the spatial integrals and the momentum integrals for $\psi^{\dagger}_n$ and $\psi_n$).  The integrals over $y$ and $z$ yield delta functions which reduce the total number of integrals to five.  These integrals are easier to compute numerically if you limit the range of values for $x$, noting that large values make small contributions to the total magnetic moment.}) the total magnetic moment as a function of $n$.  This is plotted in figure \ref{moments}.  For small values of $n$, the total magnetic moment approaches its ordinary value, \eqref{bohrmagneton}.  However, for large values of $n$ the total magnetic moment approaches $0$.

To better understand how the magnetic moment changes with $n$, let us consider an expansion of the current density, \eqref{diraccurrentdensity},\footnote{See Refs.\ [\onlinecite[eq.\ 29]{howelectronsspin}; \onlinecite{gordon1928}; \onlinecite[pg.\ 321--322]{frenkel}; \onlinecite[pg.\ 479]{huang1952}; \onlinecite[pg.\ 504]{ohanian}].}
 \begin{align}
\vec{J} &= \frac{i e\hbar}{2 m}\left\{\psi^\dagger \gamma^0 \vec{\nabla} \psi - (\vec{\nabla} \psi^\dagger) \gamma^0\psi\right\}\underline{ - \frac{e\hbar}{2 m} \vec{\nabla}\times(\psi^\dagger \gamma^0 \vec{\sigma}\psi)}
\nonumber
\\
&\quad\ +\frac{i e\hbar}{2 m c}\frac{\partial}{\partial t}(\psi^\dagger \vec{\gamma} \psi)\ ,
\label{currentexpansion}
 \end{align}
where $\vec{\sigma}$ can be defined in terms of the familiar Pauli spin matrices as
\begin{equation}
\vec{\sigma}=\left(\begin{matrix}\vec{\sigma}_p & 0 \\ 0 & \vec{\sigma}_p \end{matrix}\right)\ .
\end{equation}
If the underlined second term in \eqref{currentexpansion} were the only contribution to the current, the magnetic moment of the electron---calculated via \eqref{totalmagneticmoment}---would be
 \begin{equation}
\int \! d^3 x \; \frac{-e\hbar}{2 m c} \psi^\dagger \gamma^0 \vec{\sigma}\psi
\ ,
\label{spinmagneticmoment}
\end{equation}
where the integrand is a magnetic moment density.\footnote{See Refs.\ [\onlinecite[eq.\ 30]{howelectronsspin}; \onlinecite[eq.\ 23]{ohanian}].}  Let us call \eqref{spinmagneticmoment} the ``spin magnetic moment'' to contrast it with the total magnetic moment in \eqref{totalmagneticmoment}.  For the states in \eqref{generalstate}, the spin magnetic moment, \eqref{spinmagneticmoment}, can be calculated as an integral in momentum space,
 \begin{equation}
\int \! d^3 p \; \frac{-e\hbar}{2 m c} |f_n(\vec{p}\,)|^2 u^{\dagger}_1(\vec{p}\,) \gamma^0 \vec{\sigma} u_1(\vec{p}\,)
\ .
\end{equation}
For small values of $n$, the other contributions to the total magnetic moment are negligible and the spin magnetic moment is approximately equal to the total magnetic moment.  As $n$ increases, the spin magnetic moment and the total magnetic moment begin to differ.  The spin magnetic moment drops to two thirds of its original value as the total magnetic moment drops to $0$ (see figure \ref{moments}).\footnote{In addition to the small electron states studied here, there are many other electron states where the total magnetic moment is not the Bohr magneton.  Consider, for example, an equal superposition of a $z$-spin-up wave packet centered at one location and a $z$-spin-down wave packet centered at another location, which will have a total magnetic moment, \eqref{totalmagneticmoment}, close to $0$ [and a total angular momentum, \eqref{totalangularmomentum}, close to $0$].  This kind of state can form after an $x$-spin-up electron passes through the inhomogeneous magnetic field of a Stern-Gerlach experiment \cite{electronsspinmeasurement}.}

We have just seen that the electron's magnetic moment can be explained without faster-than-light rotation because the magnetic moment drops close to $0$ for small electron states.  The resolution of the problem regarding angular momentum is quite different.  The total angular momentum can be calculated via
\begin{equation}
\vec{L}=\int \! d^3 x \left(\vec{x} \times \vec{G}\,\right)
\ ,
\label{totalangularmomentum}
\end{equation}
where $\vec{G}$ is the momentum density.  We can write the momentum density as
\begin{equation}
\vec{G}= \frac{1}{c^2}\rho^{\mathcal{E}} \vec{v}
\ ,
\label{energyvelocity}
\end{equation}
the relativistic continuum analogue of $\vec{p}= m \vec{v}$.  Here $\rho^{\mathcal{E}}$ is the energy density (which is $c^2$ times the relativistic mass density) and $\vec{v}$ is a velocity associated with the flow of energy.  In general, $\vec{v}$ in \eqref{energyvelocity} and $\vec{v}^{\,q}$ in \eqref{chargevelocity} will not be the same \cite{howelectronsspin}.

\begin{figure}[t]
\includegraphics[width=8.6 cm]{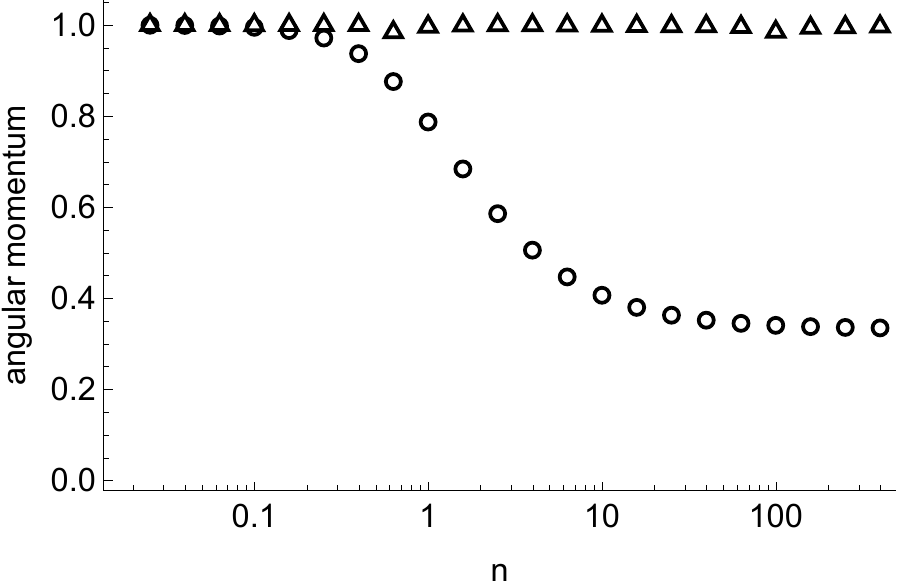}
\caption{This plot shows the total angular momentum (triangles) and the spin contribution to the angular momentum (circles) for different values of $n$, using a log scale for $n$.  The angular momentum is given as a fraction of $\frac{\hbar}{2}$.  For small values of $n$, the total angular momentum is $\frac{\hbar}{2}$ and the spin contribution is close to $\frac{\hbar}{2}$.  For large values of $n$, the total angular momentum remains constant (though there is some noise from numerical integration in the plot) and the spin contribution drops to one third of its original value.}
\label{angularmomenta}
\end{figure}

The challenge of attaining an angular momentum of $\frac{\hbar}{2}$ from \eqref{totalangularmomentum} for small electron states without faster-than-light rotation is structurally similar to the problem of attaining a magnetic moment of $\frac{e\hbar}{2 m c}$ from \eqref{totalmagneticmoment}, except for one crucial difference.  For the magnetic moment, there is a fixed amount of charge and a speed limit for the motion of that charge which together make it impossible for the electron to achieve its usual magnetic moment if its charge distribution is too compact.  For the angular momentum, the amount of energy is not fixed and thus even with a speed limit on the velocity in \eqref{energyvelocity} and a tightly peaked distribution of energy, it is possible for the electron to have a large angular momentum.

Let us consider the densities of energy and momentum that appear in the symmetrized energy-momentum tensor,\footnote{See Refs.\ [\onlinecite{howelectronsspin}; \onlinecite[appendix 7]{heitler}; \onlinecite{ohanian}; \onlinecite[sec.\ 20]{wentzel}].}
\begin{align}
\rho^{\mathcal{E}}&=\frac{i \hbar}{2}\left(\psi^\dagger\frac{\partial \psi}{\partial t}-\frac{\partial \psi^\dagger}{\partial t}\psi\right)
\nonumber
\\
&=m c^2 \psi^\dagger\gamma^0\psi + \frac{\hbar c}{2i}\left[\psi^\dagger\gamma^0\vec{\gamma}\cdot\vec{\nabla}\psi-(\vec{\nabla} \psi^\dagger)\cdot\gamma^0\vec{\gamma}\psi\right] 
\label{energydensity}
\\
\vec{G}&=\frac{\hbar}{2 i}\left[\psi^\dagger \vec{\nabla} \psi - (\vec{\nabla} \psi^\dagger)\psi \right]\underline{+\frac{\hbar}{4}\vec{\nabla}\times(\psi^\dagger \vec{\sigma} \psi)}\ .
\label{momentumdensity}
\end{align}
The move from the first line of \eqref{energydensity} to the second uses the free Dirac equation \eqref{dirac}.  For the electron states in \eqref{generalstate}, the total energy found by integrating \eqref{energydensity} differs depending on $n$.  As $n$ approaches $0$, the $m c^2 \psi^\dagger\gamma^0\psi$ term in \eqref{energydensity} dominates and the total energy approaches the standard rest energy of the electron, $m c^2$.  As $n$ increases, the contribution to the energy from the $m c^2 \psi^\dagger\gamma^0\psi$ term drops to approach $0$ and the contribution from the remainder of \eqref{energydensity} grows without bound.  These electron states can have an arbitrarily high energy.

In \eqref{momentumdensity}, there is an underlined spin contribution corresponding to an angular momentum of
\begin{equation}
\int \! d^3 x \; \frac{\hbar}{2}\psi^\dagger \vec{\sigma} \psi
\ ,
\label{spinangularmomentum}
\end{equation}
where the integrand is an angular momentum density.  We can compute the total and spin angular momenta for the states in \eqref{generalstate} using numerical integration.  For small values of $n$, the spin angular momentum \eqref{spinangularmomentum} is the main contribution to the total angular momentum \eqref{totalangularmomentum}.  As $n$ increases, the spin angular momentum drops to a third of its original value but the other contribution to the total angular momentum grows to compensate so that the total angular momentum remains constant at $\frac{\hbar}{2}$ (see figure \ref{angularmomenta}).

Although the focus in this section has been on the bare electron and not the electromagnetic field that surrounds it, one could study the way in which the charge density and current density in \eqref{diracchargedensity} and \eqref{diraccurrentdensity} act as source terms for the electromagnetic field.\footnote{Going further, one could move beyond the free Dirac equation and consider the effect of the electromagnetic field produced by an electron on that electron.  This raises problems of self-interaction that are beyond the scope of this paper (see \cite[sec.\ 8]{positrons} and references therein).}  For small electron states, one would find that the surrounding electromagnetic field becomes very strong near the electron and that the total energy in the electromagnetic field is very large.  This would yield an increase in the energy of the dressed electron for small electron states that is distinct from the increase in the energy of the bare electron just discussed in relation to \eqref{energydensity}.  Also, the electromagnetic field around the electron will carry angular momentum that should be considered when calculating the total angular momentum of the dressed electron.

\section{Proving a Minimum Size}

In his doctoral dissertation, \textcite[sec.\ 2.3]{chuu2010thesis} gives an explicit and detailed proof that there is a minimum size for wave packets composed of positive-frequency solutions to the Dirac equation.  \textcite{chuu2010} present a brief summary of the proof.  Given the counterexample states in the previous section, the existence of such a proof is puzzling.  Their proof hinges on the assumption that the wave packet has a small spread  in momentum space, which is part of their ``semiclassical'' treatment.  This assumption is stated explicitly in the dissertation, Ref.\ \cite[pg.\ 11, 14]{chuu2010thesis}, and in an arXiv preprint \cite{chuu2007} but not in the published article  \cite{chuu2010}.  In this section, we follow these authors in using quantum notation and terminology (though their result would apply just as well to the classical theory of the Dirac field mentioned in section \ref{intro}).

Chuu \textit{et al.\ }introduce a Hermitian operator $\widehat{P}$ that projects onto positive-frequency modes and a Hermitian operator $\widehat{Q}$ that projects onto negative-frequency modes.  For these projection operators, $\widehat{P}^2 = \widehat{P}$, $\widehat{Q}^2 = \widehat{Q}$, and $\widehat{P}+\widehat{Q}=\widehat{I}$.  Let us consider a purely positive-frequency state, $|\psi\rangle$, such that $\widehat{P}|\psi\rangle=|\psi\rangle$.  Let us further assume that this state is centered at the origin, $\langle \vec{x}\, \rangle = 0$.  The mean square radius, \eqref{chargeradius}, for such a state is
\begin{align}
\langle |\vec{x}|^2 \rangle &= \langle \vec{x} \cdot (\widehat{P}+\widehat{Q}) \vec{x} \, \rangle
\nonumber
\\
&= \langle \vec{x} \cdot \widehat{P} \vec{x}\,\rangle + \langle \vec{x} \cdot \widehat{Q} \vec{x} \, \rangle
\ .
\label{projected}
\end{align}
We can write the first term as
\begin{equation}
\langle \vec{x} \cdot \widehat{P} \vec{x} \, \rangle = \langle \psi | \big(\widehat{P} \vec{x} \widehat{P}\,\big)^{\dagger} \cdot \widehat{P} \vec{x} \widehat{P} | \psi \rangle
\ ,
\label{part1}
\end{equation}
which must be greater than or equal to $0$ because it is the sum of the norms of $\widehat{P} x \widehat{P}| \psi \rangle$, $\widehat{P} y \widehat{P}| \psi \rangle$, and $\widehat{P} z \widehat{P}| \psi \rangle$.  Given that the first term in \eqref{projected} is not negative, if we can derive a minimum value for the second term we will have derived a minimum size for electron states.

Although Chuu \textit{et al.\ }consider more general states in their proof, to see more easily where the key assumption comes in let us focus on the counterexample states in \eqref{generalstate}.  \textcite{chuu2010thesis} derives an expression for $\langle \vec{x} \cdot \widehat{Q} \vec{x} \, \rangle$ which can be written using \eqref{generalstate} as\footnote{The expression in \eqref{part2} can be derived from the third line of Chuu's \cite{chuu2010thesis} equation 2.23 using the familiar definitions of the Pauli spin matrices and Chuu's definition of the ``velocity operator of interband transition between positive and negative energy states'' (in the terminology of \textcite{feshbach1958}, this is a piece of the odd part of the velocity operator; on page 39 of their article, Feshbach and Villars call the odd part of the velocity operator ``the velocity operator of the Zitterbewegung'').  Note that Chuu's $\vec{q}$ differs from our $\vec{p}$ by a factor of $\hbar$.}
\begin{align}
&\langle \vec{x} \cdot \widehat{Q} \vec{x} \, \rangle = \int \! d^3 p \; |f_n(\vec{p}\,)|^2 \left(\frac{\hbar c}{2 E(\vec{p}\,)}\right)^2 \times
\nonumber
\\
&\quad\left(3 - \frac{2 |\vec{p}\,|^2 c^2}{E(\vec{p}\,)(E(\vec{p}\,)+m c^2)} + \frac{|\vec{p}\,|^4 c^4}{(E(\vec{p}\,)(E(\vec{p}\,)+m c^2))^2} \right)
\ .
\label{part2}
\end{align}
We can evaluate this contribution to the mean square charge radius, \eqref{projected}, using numerical methods to approximate the integral.  As $n$ approaches $0$, the value of \eqref{part2} approaches $\frac{3\hbar^2}{4 m^2 c^2}$.\footnote{Chuu \emph{et al.}\ say that for this kind of narrow wave packet in momentum space (centered at the origin of momentum space), $\langle \vec{x} \cdot \widehat{Q} \vec{x} \, \rangle$ would be $\frac{\hbar^2}{4 m^2 c^2}$, which differs from the above expression by a factor of $3$ [\onlinecite[pg.\ 534]{chuu2010}; \onlinecite[sec.\ 2.3]{chuu2010thesis}].  I trace the disagreement to an error in Ref.\ \cite{chuu2010thesis}---even with the assumption of a small momentum spread, the fourth line of Chuu's equation 2.23 does not follow from the third.}  However, for large values of $n$ \eqref{part2} becomes arbitrarily small.  Chuu \emph{et al.}\ claim to derive a minimum mean square charge radius for electron states by showing that $\langle \vec{x} \cdot \widehat{Q} \vec{x} \, \rangle$ has a minimum value on the order of the Compton radius squared.  But, as we have just seen, the value of this quantity can dip below such a minimum value when we relax their assumption of a narrow spread in momentum space.  In general, there is no minimum size for wave packets composed of positive-frequency modes.

\section{Conclusion}

Thus far, we have been asking about the possibility of small electron states simultaneously within two distinct physical theories: single-particle relativistic quantum mechanics and classical Dirac field theory.  However, if our goal is to understand real electrons in nature, we have a better option: quantum electrodynamics, a quantum field theory.  In this final section, I will briefly explain what I take to be the relevance of the analysis in this paper for our understanding of quantum field theory.  As there is controversy concerning the foundations of quantum field theory, others may take different lessons.

Quantum field theories can be understood as theories that describe classical fields in quantum superpositions \cite{floreanini1988, jackiw1990, hatfield}.  For electrons, the relevant field is the Dirac field.  In \cite{howelectronsspin}, I argued that electrons really do rotate in classical Dirac field theory and responded to obstacles facing the idea that electron spin is true rotation.  One problem is that if electrons are smaller than a certain minimum size, then (given the prohibition on faster-than-light motion) there appears to be no way for their rotation to generate the correct magnetic moment and angular momentum (see section \ref{SESsection}).  In \cite{howelectronsspin}, I gave a two-part response to this problem: (a) the way that the velocity of charge flow is defined from \eqref{chargevelocity}, it cannot exceed $c$; and (b) the small sub-Compton-radius electron states that raise concerns about explaining the electron's magnetic moment and angular momentum without faster-than-light rotation are not possible.  In support of (b), I relied on the false folk theorem presented by the authors quoted in section \ref{CMSsection}: there is a minimum size, on the order of the Compton radius, for states composed of positive-frequency solutions to the free Dirac equation.  (I defended the focus on positive-frequency modes in \cite{positrons}.)  Although (b) is false, it is still possible to understand the electron's magnetic moment and angular momentum as resulting from the actual rotation of charge and energy in classical Dirac field theory.  There are other ways to respond to the problem.  Focusing on the motion of charge, we may still rely on (a) and we can also observe that the magnetic moment will drop below its usual value for small electron states---as it must, given (a).  Focusing on the motion of energy, we have seen that small electron states have large energies and thus that the angular momentum of the electron can be understood as resulting from the flow of energy without any need for faster-than-light energy flow.

Studying classical Dirac field theory teaches us about the classical states that enter quantum superpositions in quantum field theory.  Here we have seen that these classical states describe electrons whose magnetic moments and angular momenta are the result of actual rotation.  We have also seen that these classical states can describe electrons with very tightly peaked distributions of energy and charge.  It is a further question---and not one I intend to address here---whether small electron states are possible in quantum field theory.  To answer that question, one would have to analyze superpositions of classical field states.

In addition to classical Dirac field theory, we have also (at the same time) been asking about the possibility of small electron states in relativistic single-particle quantum mechanics (interpreting this theory as describing the spread-out energy and charge of the electron).  I view relativistic single-particle quantum mechanics as an approximation to quantum field theory.  This approximation cannot be trusted when the possibility of additional particles being present becomes relevant.  As mentioned in section \ref{CMSsection}, the small electron states that have been discussed here are states for which the possibility of pair creation in quantum electrodynamics is relevant and thus these are states for which single-particle relativistic quantum mechanics cannot be expected to accurately approximate quantum field theory.  Still, for understanding what is possible according to relativistic single-particle quantum mechanics (viewed as a free-standing theory that may not always yield correct predictions), it is valuable to recognize that small electron states are indeed allowed.

\vfill\null
\noindent
\textbf{Acknowledgments}
I thank Logan McCarty for bringing the papers by Bracken, Flohr, and Melloy to my attention.  I also thank Jacob Barandes, Noel Swanson, and the anonymous reviewers for helpful feedback and discussion.

\end{document}